# Auricular Bioelectronic Devices for Health, Medicine, and Human-Computer Interfaces.

**William J. Tyler**
Department of Biomedical Engineering and
Center for Neuroengineering and Brain-Computer Interfaces,
University of Alabama at Birmingham, School of Engineering and
Heersink School of Medicine, Birmingham, Alabama 35294 USA

## Abstract

Recent advances in manufacturing of flexible and conformable microelectronics have opened opportunities for health monitoring and disease treatment. Other material engineering advances, such as the development of conductive, skin-like hydrogels, liquid metals, electric textiles, and piezoelectric films provide safe and comfortable means of interfacing with the human body. Together, these advances have enabled the design and engineering of bioelectronic devices with integrated multimodal sensing and stimulation capabilities to be worn nearly anywhere on the body. Of particular interest here, the external ear (auricle) offers a unique opportunity to design scalable bioelectronic devices with a high degree of usability and familiarity given the broad use of headphones. This perspective article discusses recent design and engineering advances in the development of auricular bioelectronic devices capable of physiological and biochemical sensing, cognitive monitoring, targeted neuromodulation, and control for human-computer interactions. Stemming from this scalable foundation, there will be increased growth and competition in research and engineering to advance auricular bioelectronics. This activity will lead to increased adoption of these smart headphone-style devices by patients and consumers for tracking health, treating medical conditions, and enhancing human-computer interactions.

## Introduction

Headphones are an iconic human interface. Current and forthcoming generations of headphones or auricular bioelectronics have capabilities that will fundamentally change how we approach health diagnostics, clinical intervention, and communications. The design of headphones can be traced back to the 1890's, when Earnest Mercadier developed binaural diaphragms enabling handsfree operation for telephone operators controlling switchboards (**Figure 1A**) [1]. A couple decades later Nathaniel Baldwin is credited with inventing the first audio headphones designed to enhance naval communications aboard large and noisy ships. Marking an application transition from their industrial use in communications to personal entertainment use in the audio industry, John Koss designed the first stereo headphones for listening to music in 1958. Since, electrical, mechanical, and biomedical engineering advances have enabled headphone miniaturization, microphone and biometric sensor integration, wireless connectivity, incorporation of digital signal processors (DSP) for active noise cancellation (ANC), audio filtering, amplification, spatial audio encoding, and medical device embodiments for the treatment of health conditions (**Figure 1B-D**). Today, given their global use in daily communication and digital media consumption, we recognize headphones as being deeply connected to our inner thoughts and emotions, lifestyle, and productivity [2, 3]. As such, headphones and devices intended to be worn on the ear have inspired the design and engineering of modern bioelectronic devices intended for health, medicine, and communications.

The medical device and personal electronics industries have begun to witness barriers blurred as health agencies, care providers, consumers, and patients, motivated by health monitoring, data analytics, and predictive algorithms have begun to integrate wearables into our daily lives [4-7]. Markets suffer from no shortage of low-power consumption, wireless connected, multimodal sensor integrated clinical grade and consumer health wearables measuring heart rate, heart rate variability, respiration rate, sleep and activity patterns, metabolic activity, stress levels, and oxygen levels. Over the past decade, a great race for data-driven, predictive insights afforded by modern machine learning (M/L) and artificial intelligence (AI) methods has attracted additional engineering resources around wearable electronic research and development [8, 9]. Many engineers, scientists, and members of the semiconductor and microelectronics industries have risen to challenges in wearable design, testing, packaging, and manufacturing of a bewildering array of batteries, energy harvesters, microprocessors, MEMS accelerometers, optical, electrical, and acoustic sensors fueling growth in consumer and medical wearables [10-13]

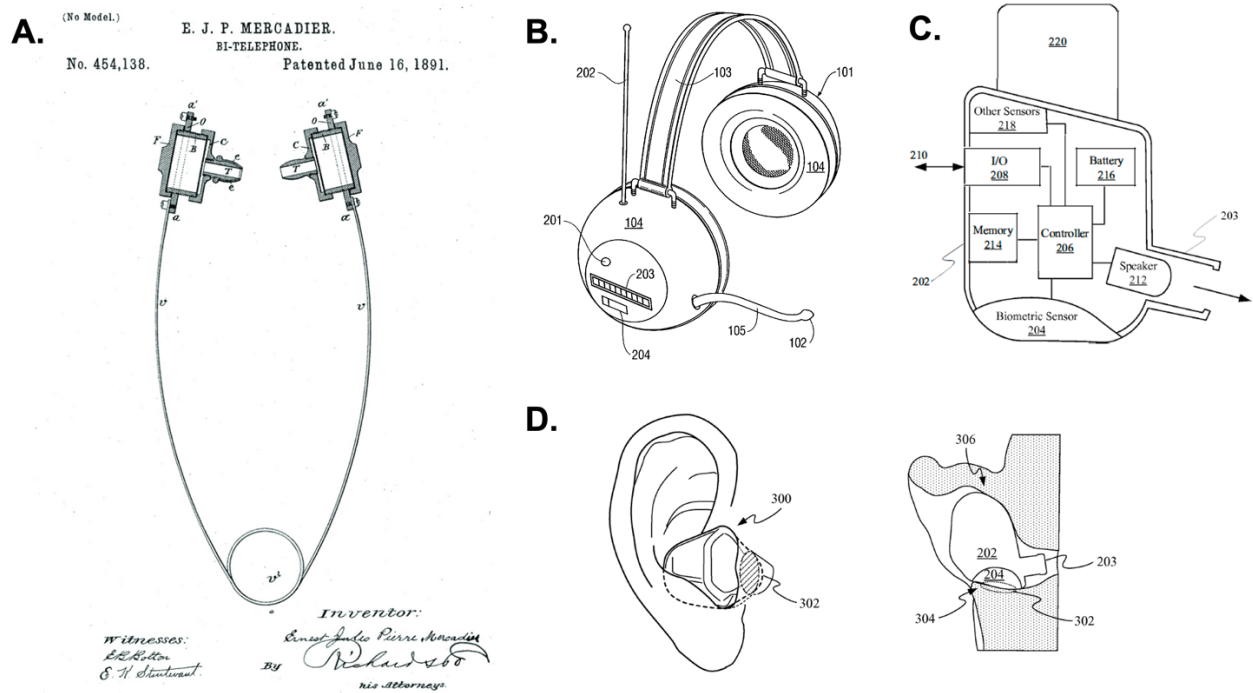

**Figure 1. Evolution of modern headphones. A.** Shown is the "Bi-Telephone" developed to enable telephone operators hands-free operation of switchboards was invented in 1891 by Ernst Mercadier and is considered to be the first modern headphone [1]. Advances in communications and microelectronics in the 1980-90's led to the development of wireless headphones. **B.** Shows a patent illustration for an embodiment of wireless headphones from the Sony Corporation [14]. Numerous advances in headphone design came about with engineering progress in wireless communication, sensor design, microelectronic packaging, digital signal processing (DSP), and automatic noise cancellation (ANC) in the 2000's. **C.** These advances led to the development of wireless, wearable headphones with integrated microphone arrays, biometric sensors, control electronics, and power management systems as illustrated by the block diagram showing components of an earbud headphone from an Apple, Inc. patent covering headphones with biometric sensing [15]. **D.** Patent illustrations showing an embodiment of Apple, Inc. earbud headphones (AirPods) with biometric sensors in the ear [15].

Developing wearable electronics that have a size, weight, and power profile sufficient for wearing as headphones has some unique engineering challenges. Physical acoustic constrains combined with the fact that people have ears of different sizes and shapes presents other challenges [16-18]. However, the consumer audio, cochlear implant, and hearing aid industries have solved many of these challenges over the past couple decades [19-22]. In fact, the FDA has recently cleared several over-the-counter (OTC) hearing aids that are now widely available to patients without the need to be fitted by an audiologist [22-25]. More recently in another innovative step forward, the FDA cleared Apple's AirPods Pro 2 consumer headphones as an OTC hearing aid software that was cleared through a *de novo* Software as a Medical Device (SaMD) regulatory pathway [26, 27].

Other advances in flexible and conformable electronics, as well as polymer materials for skin-device interfaces have enabled the development of sophisticated auricular bioelectronics embodied as headphones capable of sensing biochemical and physiological activity (**Figure 1C, D**) [15, 28, 29]. The fields of neuromodulation and bioelectronic medicine have meanwhile been developing various methods of stimulating auricular branches of cranial and cervical nerves for various outcomes [30-32]. For example, various vibrotactile and electrical forms of transcutaneous auricular vagus nerve stimulation (taVNS) have been shown to reduce inflammation including neuroinflammation associated with long COVID, reduce stress, improve sleep, decrease depression and anxiety, and enhance learning, cognition, and neurorehabilitation as further discussed below. Collectively these advances have produced a climate where the development of open- and closed-loop auricular bioelectronics will produce a new generation of medical devices, health and performance wearables, and brain-computer interfaces (BCIs) that are as scalable as personal headphones. The goal of this perspective is to highlight the anatomy, physiology, and recent engineering milestones enabling the development of modern auricular bioelectronics.

## Neural, Vascular, and Lymphatic Anatomy of the External Ear

The structure of the auricle or external ear serves mammals unique physiological roles and is tied closely to our evolution and survival [33, 34]. The external ear is a cartilaginous structure that has dense vasculature and sensorimotor nerve innervation to help foster heat dissipation, sound location, and positional awareness. The structure of the external ear has two essential parts, which are the pinna and the external auditory meatus (EAM) or outer ear canal ending at the tympanic membrane where the middle ear begins (**Figure 2A**). This location provides proximal access to the brain, which is useful for recording brain activity as discussed below. The pinna has many distinct anatomical features, which filter and direct sounds to the EAM as illustrated in **Figure 2A**, The EAM directs and conducts these sound waves to the middle ear that in turn transmit them to the inner ear for the initial phases of auditory sensory transduction and processing.

The external ear is innervated by two sensory cervical (C2-3) brachial nerves known as the great auricular nerve (GAN) and the lesser occipital nerve (LON; **Figure 2B**). It is also innervated by three different sensory cranial nerves (CN), which are the eighth CN nerve or facial nerve (CN VIII), branches of the third branch of the fifth CN nerve (CN V3) via the auriculotemporal nerve (ATN), and auricular branches of the vagus nerve (ABVN) originating from the superior ganglion of the tenth CN nerve or vagus (CN X). These sensory nerves have overlapping distributions throughout the skin of the ear differentially innervating the helix, concha, fossa, tragus, lobule, EAM, and other regions (**Figure 2B**). They also serve common functions, such as mediating physiological reflexes and gating neurophysiological arousal via the ascending reticular activating system [35, 36]. For example, ABVN innervation of the EAM serves the anatomical basis for Arnold's cough reflex [37-39]. The EAM is also innervated by branches of the ATN and facial nerve (**Figure 2B**), which reflects their close functional relationship in underlying the mammalian diving reflex [40-43] and trigemino-cardiac reflexes [44-46]. Stimulation of the ABVN can also trigger anti-inflammatory responses by modulating cytokine activity [47-50]. Given these and other effects discussed below, methods and devices for stimulation of peripheral nerves of the external ear have broad biomedical utility (see below, *Methods and Effects of Auricular Neuromodulation*).

The vascular structure of the external ear includes a rich arterial supply, as well as dense venous and lymphatic drainage system. Several types of perivascular sympathetic and parasympathetic nerve fibers regulate the vasomotor activity of the ear [51]. The external ear receives its primary blood supply from the superior anterior auricular artery, a branch of the external carotid artery. This artery is essential for providing blood to most parts of the external ear, including the anterior and inferior aspects (**Figure 2C, D**). Additionally, the superior auricular artery serves as a connection between the superficial temporal artery and the middle anterior auricular artery, ensuring a robust collateral blood flow to the ear in case of reduced blood supply from one source. The anterior auricular branch of the superficial temporal artery specifically supplies the anterior portion of the external ear. Further vascularization comes from the occipital artery, which contributes blood to the posterior region of the external ear. The EAM receives blood from both the inferior auricular artery and the auricular branches of the maxillary and superficial temporal arteries (**Figure 2C, D**). Venous drainage closely follows the arterial supply, with the veins of the external ear running alongside the corresponding arteries. This arteriovenous network is believed to play a major role in thermoregulation. Venous flow from the EAM drains into the pterygoid plexus, external jugular vein, and maxillary vein.

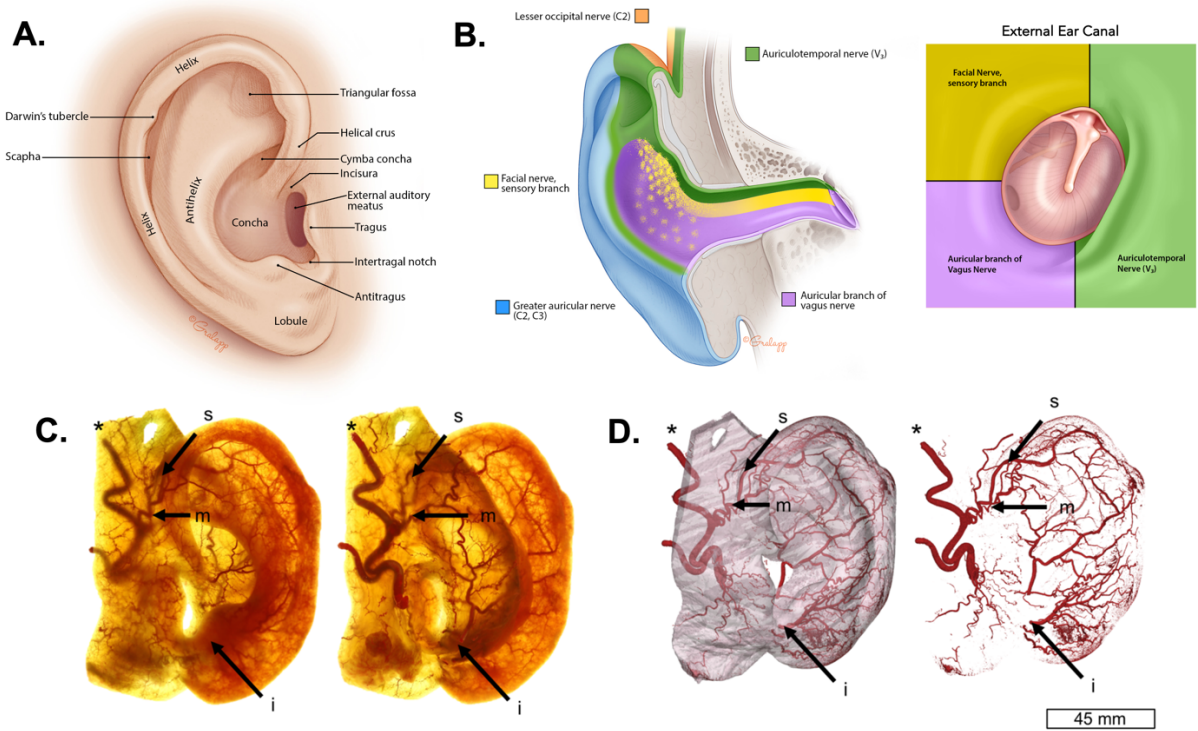

**Figure 2. Anatomy of the external ear. A.** The anatomical illustration depicts the anatomy of the human external ear showing prominent structures. **B.** The illustrations depict the sensory innervation of the external ear by the lesser occipital nerve (C2; *orange*), the great auricular nerve (C2, C3; *blue*), the facial nerve (*yellow*), the auriculotemporal nerve (ATN) or third branch of the trigeminal nerve (V$_3$; *green*), and the auricular branch of the vagus nerve (ABVN; *purple*). The image on the *right* shows this innervation in the external auditory meatus (EAM) [52]. **C.** The images show the lateral (*left*) and posterior (*right*) view of a Spalteholz ear with the auricular vasculature including the superior anterior auricular artery (s), middle anterior auricular artery (m), inferior anterior auricular artery (i), and the superficial temporal artery* [51]. **D.** The images show three-dimensional reconstructions produced using micro computed tomography (µCT) of auricular vasculature of the external ear (*left*) and for isolated auricular vascularity (*right*) [51]. The images in panels **A** and **B** were reproduced with permission from the illustrator Chris Gralapp [52]. The images in panels **C** and **D** were modified from reference [51].

Lymphatic drainage of the external ear is achieved through four primary lymphatic vessels [53]. The anterior region of the ear is drained by lymphatic branches that converge into a single vessel, which then flows into the pre-auricular lymph nodes. Lymphatic vessels from the superior aspect of the helix may travel along the anterior part of the ear, ultimately reaching the infra-auricular lymph nodes. Similarly, vessels originating in the scaphoid fossa, adjacent to the auricular tubercle, form the middle branches that drain the anterior region of the external ear, also converging toward the infra-auricular nodes. Finally, the lobular branches begin as a network of vessels in the auricular lobule. These vessels merge and drain toward the infra-auricular lymph nodes, forming an organized system of lymphatic drainage across the external ear. The unique anatomy and physiology of the external ear, including its rich vasculature, diverse sensory innervation, and proximity to the brain, make it an ideal location for recording physiological, biochemical, and brain activity data, while also providing an accessible site for stimulating nervous system activity as discussed below.

## Advances in Materials Engineering for Bioelectronic Devices

A major challenge for bioelectronic devices arises from material incompatibility between hard charge carriers (i.e., metal electrodes) and the soft, irregular surface of the skin. Conductive hydrogels, polymers, and biomedical adhesives have been developed to possess skin-like mechanical and electrical properties thereby mitigating many issues encountered when interfacing bioelectronic devices with the body [10, 54-57]. Hydrogels and conductive polymers have proven particularly advantageous in neuroengineering through the development of soft bioelectronics for neural sensing and neuromodulation interfaces [58-60]. Due to their low melting points and other physicochemical properties, liquid metals for the engineering of flexible, stretchable, and wearable electronics have opened new possibilities in bioelectronics [61]. Liquid metals printed onto or incorporated into different hydrogel and polymer substrates have enabled the development of electric skin, tattooable circuits, neural interfaces, electronic vessels, and soft thermoelectric heaters [10, 12, 61-63].

Coating, spinning, or impregnating natural and synthetic fibers with electric inks and conductive polymers has led to the development of electronically active, smart textiles and fabrics for wearable bioelectronics [64, 65]. These innovations in E-textiles have improved the comfort of devices since conducting metal fibers historically used create uncomfortable sensations against the skin despite their superior electrical conductivity. Carbonized nanoparticles and nanocomposites are presenting interesting approaches to the development of wearable electronic textiles. For instance, chemical vapor deposition of graphene monolayers and transfer from copper substrates has been used to manufacture transparent, flexible graphene fibers capable of serving as textile electrodes [66, 67]. Thin piezoelectric films and electroactive papers provide methods of fabricating flexible and conformable mechanically active sensors and actuators, as well as means for energy harvesting and power generation [68-70]. One interesting application is their use in producing piezoelectric textiles, which pose intriguing possibilities for the future of healthcare and power generation [71, 72]. With the materials, fabrication, and engineering knowledge gained over the last decade in wearable microelectronics, nearly any anatomy can be targeted and affixed with sensors and stimulators that seamlessly integrate hardware with the body. The remainder of this perspective will focus on the human external ear as a target for bioelectronic devices and applications in healthcare, medicine, and communications.

## Methods and Effects of Auricular Neuromodulation

The activity of the vagus nerve underlies core aspects of our health including digestion, cardiovascular reflexes, cardiac activity, immune responses, arousal (sleep/wake, consciousness, and fight/flight/freeze), attention, cognition, learning, and memory. Transcutaneous auricular vagus nerve stimulation (taVNS) involves the non-invasive modulation of auricular branches of the vagus nerve (ABVN; Arnold's nerve or Adleman's nerve) located under the skin's surface of the external ear. Using pulsed electrical currents to modulate ABVN fibers located in different locations of the external ear, this approach has gained attention for its safe ability to modulate autonomic nervous system activity, inflammation, neuroplasticity, attention, stress, learning, mood, and sleep, by altering activity of brain nuclei and neurotransmitters known to regulate these processes such as the locus coeruleus and norepinephrine respectively [31, 32, 36, 73-82].

Most methods and devices employing taVNS to date however fall short in providing user comfort due to their reliance on inefficient body-electrode coupling approaches. Discomfort during taVNS manifests as an electrical biting sensation, in part due to electromechanical mismatches between the electrode and skin. High current densities produced by metal electrodes with a small surface area of skin coupling, and metal or rubber electrodes clipped onto the ear create mechanical pinching sensations that are distracting and uncomfortable (**Figure 3A**). This can be further aggravated by wet coupling methods using saline sprays or electrolyte gels where the microfluidic interface (and local impedance) undergoes frequent fluctuations and distortions between the electrode and skin. For enhancing cognition, reducing stress, or promoting sleep it is critical that the user or patient has a comfortable experience, or the off target, distracting and uncomfortable sensations can override intended taVNS outcomes [83, 84]. In other words, stimulating the external ear with electrical currents can both activate and suppress sympathetic activity (stress) depending on many variables including interface comfort and usability.

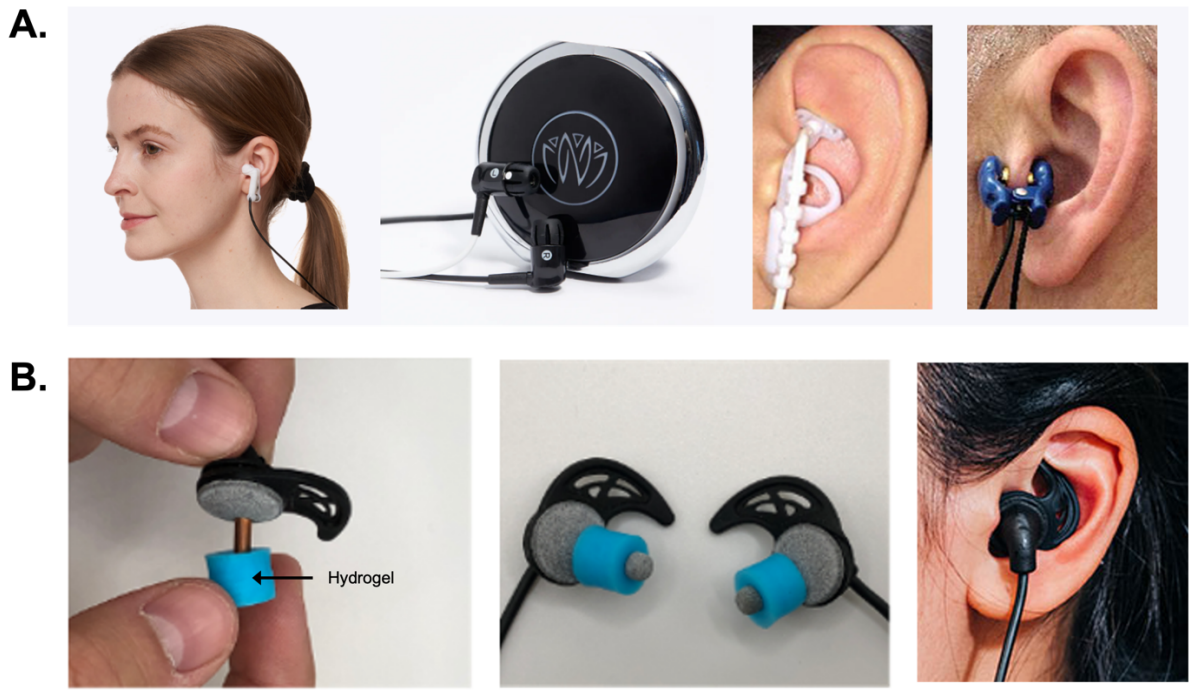

**Figure 3. Electrical methods of auricular neuromodulation. A.** Some transcutaneous auricular vagus nerve stimulation (taVNS) methods and devices utilize metal clip electrodes like the clip (Soterix Medical, Inc) shown on the *left*. These clips are used to mechanically couple the skin to a metal electrode using an electrolyte solution or gel. This approach creates a distracting pinching sensation and can produce electrical biting or prickling sensations. The Xen (Neuvana, Inc) aVNS device shown (*middle-left*) implements a different skin-electrode coupling approach using a saline-sprayed conductive rubber electrode placed in the left external acoustic meatus. This creates a distracting wet feeling in the ear of users. Due to body motion, fluid flux, absorption, and dehydration this wet coupling method also leads to electromechanical distortions in the fluid coupling layer between the charge carrier and irregular surfaces of the skin. Other taVNS methods like the Nemos device (Cerbomed GmbH) shown *middle-right* use small, steel ball electrodes that can produce high current densities resulting in discomfort or electrical biting and stinging sensations. The Tinnoff device (SaluStim Group) shown at *right* features a different type of taVNS clip electrode that can cause distracting sensations as discussed. **B.** Images of the BRAIN Buds taVNS electrodes (IST, LLC) shown in the *left* and *middle* panels were developed as conductive hydrogel earbud electrodes to produce an easy-to-use, comfortable user-experience. Using conductive hydrogels to couple electrodes to the skin results in more uniform current distributions and enhanced user comfort during transcutaneous electrical stimulation. As shown on the *right*, BRAIN Buds were designed as a bilateral taVNS system to be used like earbud headphones.

To overcome human factors issues associated with taVNS, we originally taught methods of using low-impedance, conductive polymers and soft hydrogel earbud electrodes or electrode interfaces inserted in the EAM to achieve external ear stimulation (**Figure 3B**) [85, 86]. Earbud electrodes comprised of a hydrogel inserted in this location target the ABVN, facial nerve, and auriculotemporal branches of the trigeminal nerve located just under the skin of the walls of the external acoustic meatus (**Figures 2B and 3B**). These external ear stimulation approaches, using conductive hydrogel earbud electrodes inserted into the EAM, formed the basis of methods used and devices designed to enhance foreign language learning [82, 87] and relieve tinnitus symptoms (**Figure 4A**) [88]. A major reason modern bioelectronic devices use hydrogel coupling methods is that they reduce electromechanical mismatches across skin-electrode interface resulting in stable, uniform current distributions, enhanced user comfort, and improved electrical efficiency [54-56, 89-92]. Using hydrogel earbud electrodes to achieve comfortable, bimodal, electro-aural stimulation of the external ear during presentation of notch-filtered audio stimuli was recently shown capable of reducing symptoms associated with tinnitus (**Figure 4A**) [88]. A recent study comparing different methods of VNS on language learning failed to optimize user comfort using the hydrogel coupling method previously described [82, 85]. Miyatsu and colleagues (2024) rather implemented a saline sprayed, conductive rubber earbud electrode that can be distracting and uncomfortable (**Figure 3A**), which was likely a major contributing factor to their observations that taVNS did not produce significant effects on language learning [83]. Using exact same device as Miyatsu and colleagues (2024), a previous study had shown that modifying the manufacturer earbud electrodes using a hydrogel coupling method [85] led to significant improvements in foreign language learning [82]. While other factors may have also contributed to these different observations, future efforts aimed at optimizing electroconductive hydrogels and polymers for interfacing with the ear combined with studies advancing neurostimulation algorithms and parameters will continue to enhance the usability and efficacy of taVNS for various applications. Other modes of auricular neuromodulation are also ripe for advancement.

Caloric vestibular stimulation (CVS) involves the thermal modulation of the vestibular system by cooling or heating the ear canal (**Figure 4B**). Studies have shown that CVS can modulate brain activity across different cortical regions [93], regulate mood and affect [94], reduce pain evoked potentials [95], and modulate sensory perception and conscious experience in healthy and brain-damaged patients [96-98]. A recent study using a wearable, solid-state device embodied as an aluminum, thermal headphone probe to achieve cyclic CVS was effective at reducing both motor and non-motor symptoms in Parkinson's disease patients following eight weeks of at home treatment [99]. Advances in flexible and wearable thermoelectric materials may offer new headphone design opportunities for CVS in therapeutic neuromodulation applications [99, 100]. Other materials advances in thermally conductive polymers and use of liquid metals for interfacing thermoelectric materials with the skin can improve thermal transfer and insulation to optimize CVS methods [61].

Vibrotactile and haptic stimulation of external ear has also shown to have interesting biomedical applications including to mediate human-computer interactions. Targeting vagal nerves innervating the cymba concha, vibrotactile stimulation of the external ear has been shown to modulate arousal [101] and reduce cytokine production of TNF, IL-1β and IL-6 to attenuate systemic inflammation in patients with Rheumatoid arthritis [102]. It has been argued the tactile sensitivity of the external ear has been overshadowed by its auditory functions and that haptic stimulation of the ear represents an opportunity for information transfer [103]. Lee and colleagues (2019) demonstrated small ear worn haptic stimulation devices could encode environmentally relevant spatiotemporal information by stimulating six different locations on the external ear. In an adaptive embodiment, ear haptics were demonstrated as a human-computer interface to enhance the experience of virtual reality applications for deaf and hard-of-hearing (DHH) individuals [104]. Haptic stimulation of the ear can convey sound direction in relation to DHH users during a VR experience when a system was not universally designed and intended for hearing enabled persons using spatially encoded audio to simulate sound distance [104]. The integration of piezoelectric thin films, piezopolymers, and electroactive papers [68] into flexible and conformable auricular haptic bioelectronic devices opens fascinating possibilities for medicine and communications. Not only can these materials provide for the design of active stimulation or neuromodulation devices to be worn in the ear, but they can also serve the basis for a wide range of electrophysiological and biochemical sensors to record data for diverse applications in health monitoring and cognitive enhancement.

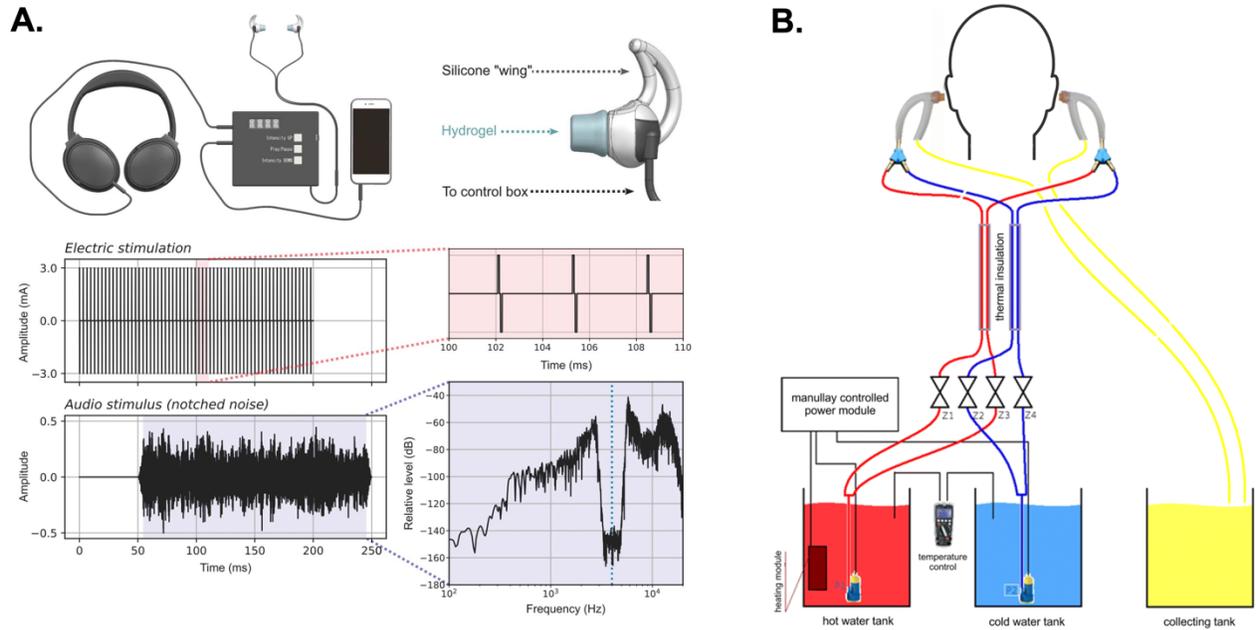

**Figure 4. Multimodal methods of auricular neuromodulation. A.** The figure illustrates a recent approach combining bilateral, pulsed electrical stimulation of the external ear using hydrogel earbud electrodes inserted into the external acoustic meatus (EAM) presented with notch-filtered auditory stimulation for the treatment of tinnitus developed by the Bose Corporation [88]. **B.** The figure illustrates a caloric vestibular stimulation (CVS) system that uses bilateral thermally conductive probes inserted into the EAM. In this embodiment the device uses either hot or cold water circulated through the ear probes to achieve CVS through heating or cooling [93]. The images in panel **A** were adapted from reference [88] and the images in panel **B** were adapted from reference [93].

### Design and Application of Auricular Sensors

In some embodiments, auricular bioelectronic devices have sensors that can detect a wide range of physiological, biometric, and chemical signals such as heart rate, brain activity using electroencephalography (EEG), head orientation, and even metabolic markers like lactate. The development of auricular monitoring devices presents challenges like those discussed above for neuromodulation. Integrating advanced sensor materials into devices intended to be worn in or on the ear while ensuring comfort, durability, and accuracy of them in real-world settings can be difficult. The development and use of hydrogels and dry EEG electrode materials represent significant advances over traditional wet electrodes, which require gels and skin preparation. Advances in these materials, flexible electronics, and additive manufacturing has improved the efficiency and comfort of auricular bioelectronic devices making them suitable for continuous everyday use.

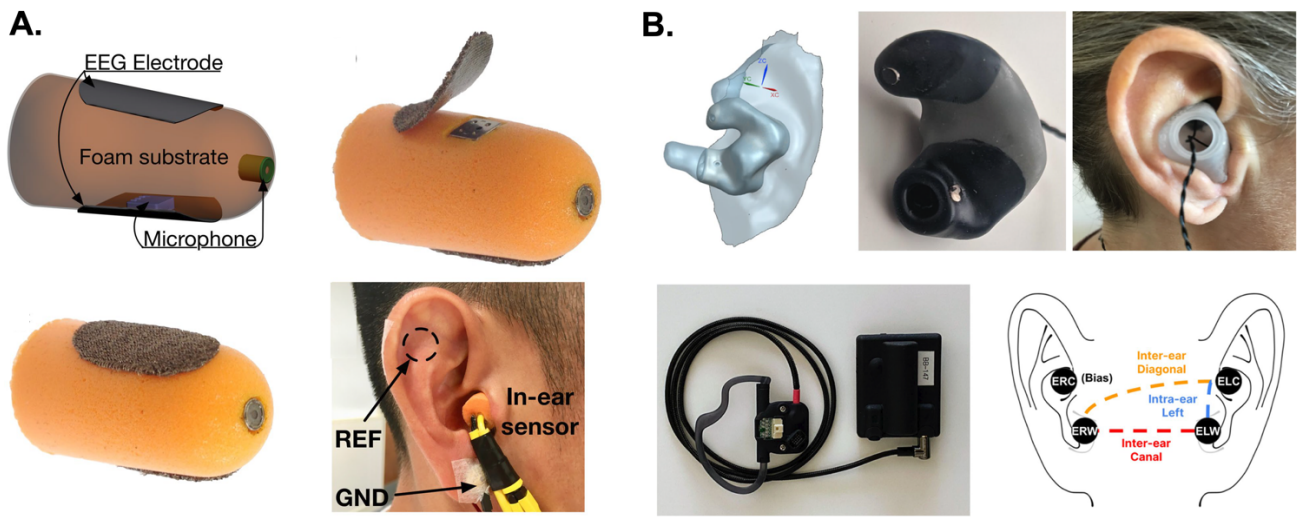

**Figure 5. Electrophysiological sensing methods for auricular bioelectronics. A.** The images depict an in-ear EEG electrode constructed using viscoelastic foam and Ag impregnated fabric used to record brain activity at rest, during evoked potentials, and during sleep [105]. **B.** Custom bilateral, auricular EEG electrodes made from a soft silicone, molded from high-resolution optical scans of the external ears of patients. The auricular EEG system was used to monitor epileptic seizure activity demonstrating good performance compared to scalp and intracranial EEG [106]. The images in panel **A** were adopted from reference [105] and the images in panel **B** were adopted from reference [106].

Unlike traditional scalp EEG systems, which rely on wet electrodes and conductive gels to reduce impedance, in-ear systems offer a more user-friendly and comfortable experience. For example, in-ear EEG electrodes can be fabricated using viscoelastic materials (memory foam) and silver-coated fabric electrodes (**Figure 5A**). This approach allows EEG sensors to fit in the EAM while capturing brain signals such as alpha rhythms, visual evoked potentials (VEPs), steady-state visual evoked potentials, and auditory steady-state responses [105]. The approach was also useful for conducting polysomnography or recording brain activity during sleep [105]. A more custom approach involves the fabrication of individualized EEG earbud electrodes. Joyner and colleagues (2024) recently used optical scans of individual patient's ears to create custom EEG earbud electrodes made from a soft silicon material and conductive polymer-coated silver rivets (**Figure 5B**). These electrodes were used to monitor epileptic activity in validation studies, which demonstrated the earbuds can provide patients with a discrete and comfortable EEG device for continuous monitoring while offering clinicians reliable and accurate data compared to intracranial and scalp recording methods [106]. Beyond this type of clinical diagnostic application, there are opportunities to develop auricular bioelectronics for brain-computer interfaces (BCI's) and human performance monitoring.

EEG metrics provide insights into how focused or mentally engaged a person may be, making them invaluable for applications in cognitive performance and mental health. By tracking these metrics, auricular bioelectronics can offer real-time feedback on mental states, enhancing both medical and consumer applications. To reach the scalable potential of these approaches however, highly personalized auricular bioelectronics as described above will need to be developed. Another innovative approach towards developing individualized, in-ear bioelectronics was recently demonstrated using spiral shaped, electrothermal actuating electrodes (SpiralE) that conform to the EAM structure of users (**Figure 6A**) [107]. Wang and colleagues (2023) engineered a flexible electrode using double-layer shape memory polymers embedded in an electrothermal actuation layer with an EEG detection top layer comprised of Au wires insulated in polyimide [107]. This design enabled in-ear EEG electrodes to comfortably conform to the shape of the ear of individual users (**Figure 6A**), while being worn and used for high fidelity recordings in visual and auditory brain-computer interfaces (BCI's) [107]. In addition to EEG monitoring, these devices have incorporated various other sensors, including accelerometers for tracking head movements and orientation, as well as biochemical sensors for detecting metabolic markers like lactate during physical activity.

Optical sensors can also integrated into headphones to record heart rate, SpO2 (blood oxygen saturation), $VO_{2max}$ and other physiological metrics. These optical sensors use photoplethysmography (PPG) technology to measure blood flow non-invasively, allowing for continuous monitoring of cardiovascular health in auricular bioelectronic devices [108-110]. Piezoelectric and MEMS-based sensors capable of detecting heart rate through pressure fluctuations in the ear have also been shown useful for cardiac monitoring [111, 112]. Advances in thermoforming techniques allow for the creation of custom and generic earpieces that can house multiple sensors while maintaining comfort and stability. Flexible and stretchable materials are used to integrate multiple sensors into a compact design, suitable for long-term wear. Xu and colleagues (2023) recently developed integrated electrochemical, chronoamperometry sensors with flexible Ag electrodes, which were 3D-printed, coated with a layer of PVA hydrogel, bonded to a flexible PCB, and mounted into an earphone assembly to simultaneously record lactate from sweat and EEG from the ear during exercise (**Figure 6B**) [113]. This multimodal sensing approach including biochemical measures from sweat may be useful in monitoring other variables beyond lactate during EEG, HR, and head position. For instance, obtaining measures of stress hormones or drug metabolites continuously to gain insights related to human performance using auricular bioelectronic devices may be particularly useful.

Another exciting area of development in auricular bioelectronics is the integration of spatial audio with sensor data, such as EEG and accelerometry. Spatial audio enhances the auditory experience by simulating how sound moves in the user's environment. With the help of accelerometers and gyroscopes, which track head movements in real-time, spatial audio can adjust sound orientation based on the user's head position, maintaining a consistent and immersive auditory experience. This data, combined with EEG monitoring of cognitive load, attention, and engagement can optimize the auditory experience for various activities, such as work, studying, relaxation, focus, or gaming. Wireless in-ear EEG systems, with multi-channel, multimodal recording capabilities will continue to expand the potential for BCIs and neurotechnology. As the capabilities of wearable auricular devices continue to evolve, the integration of both stimulation and sensing functions within a closed-loop system opens new possibilities for real-time neuromodulation, personalized health interventions, and advanced brain-computer interfaces.

**Closed-loop Auricular Bioelectronics**

Closed-loop, auricular bioelectronics possess transformative potential for neuromodulation by integrating real-time sensing and stimulation capabilities. One prominent application would be for the treatment of atrial fibrillation or arrhythmias, where in-ear sensors can monitor for abnormal heart rhythms while taVNS provides corrective feedback. The sensing of arrhythmic events using PPG or piezoelectric-based heart rate sensors embedded in the ear canal could trigger taVNS to regulate parasympathetic activity and restore normal cardiac rhythms [114-116]. Similarly, stimulation of vagal or trigeminal fibers may hold potential for treating some forms of sleep apnea and sleep-disordered breathing [117, 118]. In this embodiment, auricular bioelectronics can be engineered to sense abnormal breathing patterns to trigger responsive stimulation of trigeminal and ABVN fibers in the EAM to reduce airway resistance and restore proper respiration during sleep. Beyond cardiac and respiratory conditions, engineers should consider the development of auricular bioelectronics to sense biochemical markers of inflammation, such as cytokines, and modulate immune responses by electrically, thermally, or mechanically stimulating the ABVN to increase anti-inflammatory and/or reduce pro-inflammatory cytokine activity [47, 48, 73, 119-121]. Such a closed-loop auricular bioelectronic could offer personalized treatments for chronic auto-immune disorders like rheumatoid arthritis, psoriasis, and Chron's, as well as other immunological conditions disease where patients suffer from frequent inflammatory flares.

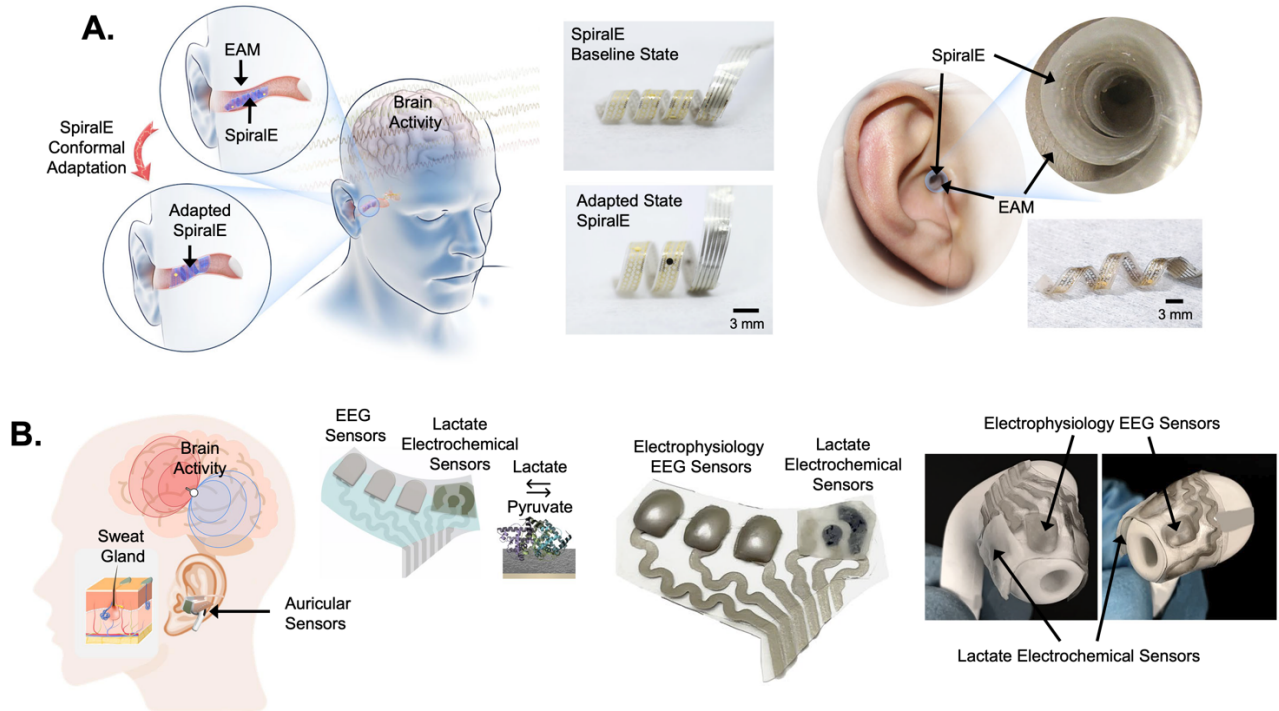

**Figure 6. Flexible auricular bioelectronics for electrophysiological and biochemical sensing. A.** The images on the *left* show a spiralized, electrothermally conforming EEG electrode (SpiralE) designed to fit in the external auditory meatus (EAM) of users [107]. The SpiralE conforms to the shape of individual user's EAM for a comfortable and electrically efficient fit. The images in the *middle* show the SpiralE in a baseline state (*top*) and adapted state (*bottom*). The images on the *right* show SpiralE inserted into the EAM of a user in its conformed state. The SpiralE auricular EEG electrode was demonstrated to be effective for recording brain activity patterns useful in auditory and visual brain-computer interface embodiments [107]. **B.** The images on the *left* depict an approach to recording brain activity using EEG sensors and recording biochemical signals from sweat glands in the ear using flexible, multimodal sensing, auricular electrodes [113]. The image in the *middle* shows a photograph of the flexible, multi-electrode array with EEG sensors and electrochemical sensors designed to detect lactate. The photographs on the *right* show the multi-electrode array mounted onto an earbud chassis to create an auricular sensor capable of sensing brain activity and lactate. This approach was demonstrated useful for recording brain activity and lactate in human subjects during exercise [113]. The images in panel **A** were adopted from reference [107] and the images in panel **B** were adopted from reference [113].

There are other promising avenues for closed-loop auricular bioelectronics, such as for the treatment of neuromuscular disorders like Restless Leg Syndrome (RLS) since taVNS has been shown to provide some benefit for the condition [122-124]. Abnormal activity recorded from wireless electromyography (EMG) sensors placed on the legs to detect episodes could be used to trigger taVNS for mitigating the discomfort and restlessness associated with RLS. In a similar fashion, EMG sensors placed in auricular bioelectronic devices could detect abnormal activity associated with nocturnal bruxism and responsively trigger taVNS stimulation protocols, which have been shown useful in reducing bruxism severity [125]. Methods for cognitive enhancement and attention regulation are ripe for development using auricular bioelectronics. In-ear EEG sensors capable of monitoring neural activity related to attention and cognitive engagement can be paired with taVNS to maintain focus or enhance vigilance during tasks that require sustained attention [77]. This has potential applications not only for those with cognitive disorders but also for enhancing human performance in high-stakes environments, such as in military or aerospace operations, where maintaining cognitive function and decision-making processes is critical [126-129].

In other human performance enhancement or health applications, closed-loop systems could help regulate stress responses by continuously monitoring physiological and biochemical stress markers. Integrated electrochemical or optical sensors in auricular devices could detect cortisol or other stress hormones, while heart rate variability sensors can report sympathetic tone. When elevated stress is detected, auricular bioelectronic devices could deliver taVNS protocols to reduce sympathetic activity and promote parasympathetic activation thereby dampening stress and improving overall well-being. Additionally, the integration of these sensors and stimulation methods with BCIs creates opportunities for enhanced human-machine interactions. As auricular bioelectronics advance, they could enable seamless, scalable BCIs that are easy to use and integrate into daily life as personal headphones are today. This would almost certainly open new frontiers for medical, cognitive, and communication technologies.

## Outlook for Auricular Bioelectronics

Auricular bioelectronics are poised to revolutionize healthcare and human performance by providing a versatile, wearable, non-invasive platform for real-time monitoring and neuromodulation. These devices, integrating advanced sensors and stimulation capabilities, hold potential for treating a wide range of conditions, from arrhythmias and sleep apnea to inflammation and cognitive disorders. As material science and flexible electronics continue to advance, auricular bioelectronics will become increasingly dynamic, comfortable, effective, and personalized. With scalable designs akin to personal headphones, these devices are likely to play a significant role in BCIs, offering yet unrealized opportunities for improving health, cognition, and human-machine interactions.

## Conclusions

While auricular bioelectronics hold great promise, several pitfalls and limitations must be addressed for these technologies to achieve their full potential. First, there are challenges associated with overcoming the technological barriers to developing reliable, long-lasting, and comfortable wearable devices. Ensuring the accuracy, reliability, durability of sensors, particularly in real-world environments with variable conditions, is essential for making closed-loop systems robust and effective. Optimizing energy efficiency will require attention since microelectronics require advanced power management systems to enable continuous operation. Several barriers to adoption might be related to the cost and complexity of engineering devices, which may limit access to those who can afford them unless manufacturing and production costs can be minimized. The lack of public understanding of potential risks, benefits, and functions of auricular bioelectronics could cause hesitancy among consumers, patients, and healthcare providers.

From a societal and ethical perspective, data privacy and security will always pose concerns and challenges. Auricular bioelectronics, which continuously collect sensitive biometric and neural data, could pose privacy risks if not properly protected and safeguarded. Another ethical consideration as alluded above is the availability and accessibility of this technology, which could exacerbate existing healthcare disparities if not distributed equitably. Despite these challenges, the next era of technological developments, including continued advances in flexible electronics, packaging, improved sensor algorithms, and improved energy harvesting methods, will drive innovation forward in the development of auricular bioelectronics. It will be critical to form collaborations between medical, research, semiconductor, and regulatory organizations to establish standards and guidelines to ensure auricular bioelectronics are safe, effective, and accessible to a broad range of users. These collaborations will ultimately pave the way for the integration of auricular bioelectronics into everyday healthcare and cognitive performance enhancement.

**Acknowledgements**
The development of BRAIN Buds by IST, LLC was funded based on research sponsored by Air Force Research Laboratory under agreement number FA8650-18-2-5402. The U.S. Government is authorized to reproduce and distribute reprints for Government purposes notwithstanding any copyright notation thereon. The views and conclusions contained herein are those of the authors and should not be interpreted as necessarily representing the official policies or endorsements, either expressed or implied, of Air Force Research Laboratory (AFRL) or the U.S. Government.

**Disclosures**
WJT is a co-founder and equity holding member of IST, LLC. WJT has several pending and issued patents related to the methods of neuromodulation described for enhancing cognition, skill training, learning, and human performance.